\documentclass{iopart}

\usepackage{graphicx}
\usepackage{dcolumn}
\usepackage{bm}

\newcommand{\Op}{\hat}
\newcommand{\beq}{\begin{equation}}
\newcommand{\eeq}{\end{equation}}
\newcommand{\beqar}{\begin{eqnarray}}
\newcommand{\eeqar}{\end{eqnarray}}
\newcommand{\bea}{\begin{eqnarray}}
\newcommand{\eea}{\end{eqnarray}}
\newcommand{\bcen}{\begin{center}}
\newcommand{\ecen}{\end{center}}

\begin{document}
\title{Irreversible performance of a quantum harmonic heat engine}
\author{Yair Rezek, Ronnie Kosloff}
\address{Department of Physical Chemistry, the Hebrew University, Jerusalem 91904, Israel}
\ead{yairr@fh.huji.ac.il}

\begin{abstract}The unavoidable irreversible loss of power in a heat engine 
is found to be of quantum origin. Following thermodynamic tradition a 
model quantum heat engine operating in an Otto cycle is analyzed, 
where the working medium is composed of an ensemble of harmonic oscillators 
and changes in volume correspond to changes in the curvature of the potential well. 
Equations of motion for quantum observables are derived for the complete cycle of operation. 
These observables are sufficient to determine the state of the system and with 
it all thermodynamical variables. 
Once the external controls are set the engine settles to a limit cycle. 
Conditions for optimal work, power, and entropy production are derived. 
At high temperatures and quasistatic operating conditions the efficiency 
at maximum power coincides with the endoreversible result $\eta_q=1-\sqrt{{T_c}/{T_h}}$. 
The optimal compression ratio varies from ${\cal C} =\sqrt{T_h/T_c}$ 
in the quasistatic limit where the irreversibility is dominated by heat 
conductance to ${\cal C} =(T_h/T_c)^{1/4}$ in the sudden limit when the 
irreversibility is dominated by friction. When the engine deviates from  
adiabatic conditions the performance is subject to friction. 
The origin of this friction can be traced to the noncommutability 
of the kinetic and potential energy of the working medium.
\end{abstract}


\maketitle

\section{Introduction}

Our cars, refrigerators, air-conditioners, lasers and power plants are all examples of heat engines. 
Practically all such heat engines operate far from the ideal maximum efficiency conditions
set by Carnot \cite{carnot}. To maximize the power output efficiency is sacrificed.
This trade-off between efficiency and power is the focus of “finite time thermodynamics". The field was initiated  
by the seminal paper of Curzon and Ahlborn \cite{curzon75}.
From everyday experience the irreversible phenomena that limits the optimal performance of engines \cite{salamon01}
can be identified as losses due to friction, heat leaks, and heat transport. 
Is there a unifying fundamental explanations for these losses? 
Is it possible to trace the origin of these phenomena to quantum mechanics? 
To address these issues we follow the tradition of thermodynamics constructing a model quantum heat engine.

Gedanken heat engines are an integral part of thermodynamical theory.
Carnot in 1824 set the stage by analyzing an ideal engine \cite{carnot,guemez}. 
Carnot's analysis preceded the systematic  
formulation that led to the first and second laws of thermodynamics \cite{callen}. 
Amazingly, thermodynamics was able to keep its independent status despite 
the development of parallel theories dealing with the same subject matter.
Quantum mechanics overlaps thermodynamics in that it describes the state of matter.
But in addition, quantum mechanics includes a comprehensive description of dynamics.
This suggests that quantum mechanics
can originate a concrete interpretation of the word {\em dynamics} in thermodynamics leading to
a fundamental basis for finite time thermodynamics.

The following questions come to mind:
\begin{itemize}
\item{How do the laws of thermodynamics emerge from quantum mechanics?}
\item{What are the quantum origins of irreversible phenomena involving friction and heat transport?} 
\item{What is the relation between the quasistatic thermodynamical process and the quantum adiabatic theorem?}
\end{itemize}
To address these issues a model of a reciprocating quantum heat engine is constructed. 
Extreme care has been taken to choose a model which can be analyzed from first principles. 
The specific heat engine chosen is a quantum version of the Otto cycle. 

The present study follows a tradition of analyzing quantum models of reciprocating 
heat engines \cite{k85,k87,k152,k176,k201,lloyd,bender00,scully03,salas04,quan2005}. Most of these models
can be classified as endoreversible. In such models the source of irreversibility
can be traced to the heat transfer between the working medium and the heat baths. 
Internal irreversibility was studied in a coupled spin working medium \cite{k176,k201}.
The present paper employs an ensemble of harmonic oscillators as a working medium
which incorporates both internal and external sources of irreversibility.
A quantum version of the Otto cycle is described in section \ref{sec:otto}.
The link between the quantum state and thermodynamical observables
is developed in section \ref{sec:thermo}.
This model allows a closed form solution of the quantum dynamics 
(section \ref{sec:otodynamics}).
At low temperature the results have similarities with the spin model, at
higher temperatures the well studied finite time thermodynamical performance 
characteristics appear naturally (section \ref{sec:engineaction}).

\section{The quantum Otto cycle}
\label{sec:otto}

Nicolaus August Otto invented a reciprocating four stroke engine in 1861
and won a gold medal in the 1867 Paris world fair \cite{otto}.
The basic components of the engine include hot and cold reservoirs, a working medium, 
and a mechanical output device.
The cycle of the engine is defined by four branches:
\begin{enumerate}
\item{The hot {\em isochore}: heat is transferred from the hot bath to the working medium without a volume change.}
\item{The power {\em adiabat}: the working medium expands producing work while isolated from the hot and cold reservoirs.}
\item{The cold {\em isochore}: heat is transferred from the working medium to the cold bath without a volume change.}
\item{The compression {\em adiabat}: the working medium is compressed consuming power while isolated from the 
hot and cold reservoirs.}
\end{enumerate}
Otto already determined that the efficiency $\eta$ of the cycle is limited to 
$\eta_o \le 1 -(\frac{V_h}{V_c})^{\frac{C_p}{C_v}-1}$ 
(where $V_{c/h}$ and $T_{c/h}$ are the volume and temperature of the working medium 
at the end of the hot and cold {\em isochore}s respectively) \cite{callen}. 
As expected, the Otto efficiency is always 
smaller than the efficiency
of the Carnot cycle $\eta_o \le \eta_c= 1  - \frac{T_c}{T_h}$.

\subsection{Quantum dynamics of the working medium}

The quantum analogue of the Otto cycle requires a dynamical description of 
the working medium, the power output and the heat transport mechanism.

The working medium is composed of an ensemble of non interacting particles described by the 
density operator $\Op \rho$. These particles are confined by an harmonic potential $V(Q)=\frac{k}{2}Q^2$.
Expansion and compression of the working medium is carried out by externally controlling the curvature of the potential
$k(t)$. The energy of an individual particle is represented by the Hamiltonian operator:
\begin{equation}
\Op H ~~=~~ \frac{1}{2m}\Op P^2 ~+ ~ \frac{k(t)}{2} {\Op Q^2}~~~,
\label{eq:hamils}
\end{equation}
where $m$ is the mass of the system and $\Op P$ and $\Op Q$ are the momentum and position
operators. All thermodynamical quantities will be intensive i.e. normalized to the number of particles.
In the macroscopic Otto engine the internal energy of the working medium
during the adiabatic expansion is inversely proportional to the volume.
In the harmonic oscillator, the energy is linear in
the frequency $\omega(t) = \sqrt{k(t)/m}$ \cite{boyer03}. This therefore plays the role of inverse volume $\frac{1}{V}$. 

The Hamiltonian (\ref{eq:hamils}) is the generator of the evolution on the adiabatic branches.
The frequency $\omega$ changes from $\omega_h$ to $\omega_c$ in a time period $\tau_{hc}$ in the power {\em adiabat} ($\omega_h > \omega_c$)
and from $\omega_c$ to $\omega_h$ in a period $\tau_{ch}$ in the compression {\em adiabat}. 
The dynamics on the state $\Op \rho$ during the adiabatic branches  
is unitary and is the solution of the Liouville von Neumann equation \cite{vNeumann}:
\begin{equation}
 \dot {\Op \rho}(t) ~~=~~ -\frac{i}{\hbar} [\Op H(t),\Op \rho(t)]~~~~,
 \label{eq:lvn} 
\end{equation}
where $\Op H$ is time dependent during the evolution. Notice that $[ \Op H (t),\Op H(t')] \neq 0$ since
the kinetic energy does not commute with the varying potential energy.

The dynamics on the hot and cold {\em isochore}s is an equilibration process 
of the working medium with a bath at temperature $T_h$ or $T_c$. This is the dynamics
of an open quantum system where the working medium is described explicitly
and the influence of the bath implicitly \cite{lindblad76,breuer}: 
\begin{equation}
 \dot {\Op \rho}(t) ~~=~~ -\frac{i}{\hbar} [\Op H,\Op \rho]+{\cal L}_D (\Op \rho)~~,
 \label{eq:lvn2} 
\end{equation}
where ${\cal L}_D$ is the dissipative
term responsible for driving the working medium to thermal equilibrium,
while the Hamiltonian $\Op H = \Op H(\omega)$ is static.
The equilibration is not complete since only a finite time $\tau_h$ or $\tau_c$ is allocated
to the hot or cold {\em isochores}. The dissipative ``superoperator'' ${\cal L}_D$ must conform to Lindblad's form for a Markovian evolution \cite{lindblad76}, and for the harmonic oscillator can be expressed as \cite{ingarden75,louisell}:
\begin{equation}
\mathcal{L}_{D}(\Op \rho)
~~=~~k_{\downarrow}(\Op a^{\dagger}\Op \rho \Op {a}~-~\frac{1}{2}\{\Op {a}\Op {a}^{\dagger},\Op \rho\})+
k_{\uparrow}(\Op{a}\Op \rho \Op {a}^{\dagger}-\frac{1}{2}\{\Op{a}^{\dagger}\Op{a},\Op \rho\})\;.
\label{eq:dissipative}
\end{equation}
where $\{ \Op A ,\Op B\}= \Op A \Op B+\Op B \Op A$. $k_{\uparrow}$ and $k_{\downarrow}$ 
are heat conductance rates obeying detailed balance  
$\frac{k_{\uparrow}}{k_{\downarrow}} = e^{-\frac{\hbar \omega}{k_b T}}$, where $T$ is either $T_h$ or
$T_c$. $\Op a^{\dagger}$ and $\Op a$ are the raising and lowering operators. 
Notice that they are different in the hot and cold {\em isochore}s since
$\Op a = \frac{1}{\sqrt 2} (\sqrt{\frac {m \omega}{\hbar}} \Op Q +i \sqrt{\frac{1 }{ \hbar m \omega}}\Op P)$
depends on $\omega$. 

Equation (\ref{eq:dissipative}) is known as a quantum Master equation. It is an example of a reduced description
where the dynamics of the working medium is sought explicitly while the baths
are described implicitly by two parameters: the heat conductivity 
$\Gamma=k_{\downarrow}-k_{\uparrow}$ and
the bath temperature $T$. The Lindblad form of (\ref{eq:dissipative}) guarantees that the density operator of the extended system (system+bath) remains positive (i.e. physical) \cite{lindblad76}. Specifically, 
for the harmonic oscillator, (\ref{eq:dissipative}) has been derived from first principles by many different authors \cite{louisell}.

To summarize, the quantum model Otto cycle is composed of a working fluid of harmonic oscillators (\ref{eq:hamils}), with volume corresponding to $1/\omega$. The power stroke is modelled by the Liouville von Neumann equation (\ref{eq:lvn}) while the heat transport via a Master equation (\ref{eq:lvn2},\ref{eq:dissipative}). It differs from the thermodynamical model in that a finite time period 
is allocated to each of these branches.
Solving these equations for different operating conditions allows to obtain the quantum thermodynamical 
observables.

\section{Quantum thermodynamics}
\label{sec:thermo}

Thermodynamics is notorious in its ability to describe a process employing an extremely 
small number of variables. A minimal set of quantum expectations 
$\langle \Op X_n \rangle$ constitutes the analogue description where 
$\langle \Op X_n \rangle = Tr \{\Op X_n \Op \rho \}$. 
 The dynamics of this set is generated by the Heisenberg equations of motion
\begin{equation}
 \dot {\Op X}  ~~=~~ \frac{\partial \Op X}{\partial t}~+~\frac{i}{\hbar} [\Op H,\Op X]+{\cal L}^*_D (\Op X)
 \label{eq:heisenberg} 
\end{equation}
where the first term addresses an explicitly time dependent set of operators, $\Op X(t)$.

To move from {\em dynamics} to {\em thermo}dynamics we need to consider the three laws. The energy expectation $E$ is obtained when $\Op X = \Op H$, i.e.  $E = \langle \Op H \rangle $.
The quantum analogue of the first law of thermodynamics \cite{sphon,k24}
$d E = d {\cal W}+ d {\cal Q}$ is obtained by inserting  $\Op H $ into (\ref{eq:heisenberg}):
\begin{equation}
 \dot {\Op E} ~~=~~ \dot{ \cal W} ~+~ \dot {\cal Q }
 ~~=~~\langle~\frac{\partial \Op H}{\partial t}~\rangle~~+~~ \langle~{\cal L}^*_D (\Op H) ~\rangle~~~.
 \label{eq:firstlaw} 
\end{equation}
The power is identified as ${\cal P} ~=~ \dot{ \cal W}~=~ \langle~\frac{\partial \Op H}{\partial t}~\rangle$. 
The heat exchange rate becomes 
$\dot {\cal Q}=\langle~{\cal L}^*_D (\Op H) ~\rangle$. 
The Otto cycle contains the simplification that power is produced or consumed
only on the {\em adiabat}s and heat transfer takes place only on the {\em isochore}s. 

The third law is trivially manifest for a system at thermal equilibrium (Gibbs state) 
at temperature zero. The intricacies of proving the second law in the general case 
are beyond this paper. Instead, in accordance with thermodynamic tradition, 
its applicability will be demonstrated for a closed cycle. 
Note that thermodynamics will be maintained even though the system 
is {\em not} generally in thermal equilibrium.

The thermodynamic state of a system is fully determined by the thermodynamical 
variables. Statistical thermodynamics adds the prescription that the state 
is determined by the maximum entropy condition
subject to the constraints set by the thermodynamical observables 
\cite{janes57a,janes57b,katz67}. Maximizing the Von Neumann entropy \cite{vNeumann}
\begin{equation}
S_{VN}=-k_B Tr\{\Op {\rho}\ln(\Op {\rho})\}
\label{eq:vnentropy}
\end{equation}
subject to the energy constraint leads to thermal equilibrium \cite{katz67},
\begin{equation}
\Op \rho_{eq}~~=~~ \frac{1}{Z} e^{-\frac{\Op H}{k_B T} }
\label{eq:req}
\end{equation}
where $k_B$ is Boltzmann constant and $Z=Tr\{e^{-\frac{\Op H}{k_B T}} \}$ is the partition function. 

The state of the working medium in general is not in thermal equilibrium. 
In order to generalize the canonical form (\ref{eq:req}) additional observables 
are required to define the state of the system.
The maximum entropy state subject to this set of  observables  
$\langle \Op X_j \rangle = tr \{ {\Op X_j} \Op \rho \}$ becomes
\begin{equation}
\Op \rho ~~=~~  \frac{1}{Z}\exp\left( \sum_j ~\beta_j \Op X_j  \right) ~,
\label{eq:state}
\end{equation}
where $\beta_j$ are Lagrange multipliers. 
The generalized canonical form of (\ref{eq:state}) is meaningful only if 
the state can be cast in the canonical form during the entire cycle of the engine
leading to $\beta_j =\beta_j(t)$. This requirement is called canonical invariance \cite{wise}.
A necessary condition for canonical invariance is that the set
of operators $\Op X$ in (\ref{eq:state}) is closed under the dynamics generated by the equation of motion.
If this conditions is also sufficient for canonical invariance, then the state of the system can be reconstructed
from a small number of quantum observables $\langle \Op X_j \rangle(t)$, which are the thermodynamical observables in the sense that they define the state under the maximum entropy principle.

The condition for Canonical invariance on the unitary part of the evolution taking place
on the {\em adiabat}s is as follows. If the Hamiltonian is a linear combination of the operators in the set $\Op H(t)= \sum_m h_m \Op X_m~$ ($h_m(t)$ are 
expansion coefficients),
and the set forms a closed Lie algebra  $[\Op X_j,\Op X_k]=\sum_l C^{jk}_l \Op X_l$
(where $C^{jk}_l$ is the structure factor of the Lie algebra), then the set $\Op X $
is closed under the evolution \cite{weinorman63}. In addition canonical invariance prevails \cite{Alhassid}. 

For the Otto cycle the set of the operators $\Op P^2$ , $\Op Q^2$ and $\Op D =  (\Op Q \Op P + \Op P \Op Q)$
form a closed Lie algebra. Since the Hamiltonian is a linear combination of the two first operators
of the set ($\Op P^2$ and $\Op Q^2$), canonical invariance will result on the adiabatic branches.

On the {\em isochore}s the set has to be closed also to the operation of  ${\cal L}_D$. 
The set $\Op P^2$, $\Op Q^2$ and $\Op D$
is also closed to ${\cal L}_D$ defined by (\ref{eq:dissipative}).
This condition is only necessary for canonical invariance. 
Nevertheless for the harmonic working medium and ${\cal L}_D$ defined in (\ref{eq:dissipative})
the closure is also sufficient for canonical invariance to take place (Cf. \ref{ap:canonical}).

The significance of canonical invariance is that all quantities become functions 
of a very limited set of thermodynamic quantum observables $\langle \Op X_j \rangle$.
The choice of operators $\Op X_j$ should reflect the most telling thermodynamical
variables. Explicitly for the current engine the thermodynamical variables are chosen 
to be the expectations of the following operators: 
\begin{itemize}
\item{The Hamiltonian 
$\Op H(t)~~=~\frac{1}{2 m} \Op P^2+\frac{1}{2}m \omega(t)^2 \Op Q^2$.} 
\item{The Lagrangian
$ \Op L(t) = \frac{1}{2 m} \Op P^2-\frac{1}{2}m \omega(t)^2 \Op Q^2$.}
\item{The position momentum correlation
$~~ \Op D=  (\Op Q \Op P + \Op P \Op Q) $.}
\end{itemize} 
These operators are linear combinations of the same Lie algebra as $\Op Q^2, \Op P^2 $ and $\Op D$.

To make explicit the connection between the expectation values of the closed
set of operators and the state of the system $\Op \rho$ an alternative
product form for the density operator is used. It is defined by the parameters $\beta$, $\gamma$ and $\gamma^*$:
\begin{equation}
\Op {\rho}~~=~~
\frac{1}{Z}e^{\gamma {\Op {a}}^{2}}e^{-\beta {\Op H}}e^{\gamma^{*}{\Op {a^{\dagger}}}^{2}}~~,
\label{eq:product}
\end{equation}
where $ \Op {H}=\frac{\hbar\omega}{2}({\Op a}{\Op a^{\dagger}}+{\Op a^{\dagger}}{\Op a}) $,
$\Op {D}  =  -i\hbar(\Op {a^2}-\Op {a^{\dagger}}^2)$ , 
$ \Op {L}=-\frac{\hbar\omega}{2}(\Op {a}^2 +\Op {a^{\dagger}}^2) $ and
\begin{equation}
Z~~=~~\frac{e^{\frac{\beta\hbar\omega}{2}}}{(e^{\beta\hbar\omega}-1)\sqrt{1-\frac{4\gamma \gamma^{*}}{(e^{\beta\hbar\omega}-1)^{2}}}}
~.
\end{equation}
From (\ref{eq:product}) the expectations of $\Op H $ and $  \Op a^2 $ are extracted leading to
\begin{equation}
\left\langle \Op {H}\right\rangle  ~~ = ~~ \frac{\hbar\omega(e^{2\beta\hbar\omega}-4\gamma \gamma^{*}-1)}{2((e^{\beta \hbar \omega}-1)^{2}-4\gamma \gamma^{*})}
\label{eq:expecth}
\end{equation}
and
\begin{equation}
\left\langle \Op {a^2}\right\rangle  ~~ = ~~ \frac{2\gamma^{*}}{(e^{\beta\hbar\omega}-1)^{2}-4\gamma \gamma ^{*}}
~~.
\label{eq:expeca}
\end{equation}
Equations (\ref{eq:expecth}) and (\ref{eq:expeca}) can be inverted leading to
\begin{equation}
\gamma ~~ = ~~ \frac{\hbar\omega(2\left\langle \Op {L}\right\rangle +i\omega\left\langle \Op {D}\right\rangle )}{4\left\langle \Op {L}\right\rangle ^{2}+\omega^{2}\left\langle 
\Op {D}\right\rangle ^{2}-(\hbar\omega-2\left\langle \Op {H}\right\rangle )^{2}}
\label{eq:gamainver}
\end{equation}
and
\begin{equation}
e^{\beta\hbar\omega} ~~ = 
~~ \frac{4\left\langle \Op {L}\right\rangle ^{2}-4\left\langle \Op {H}\right\rangle ^{2}
+\omega^{2}(\hbar^{2}+\left\langle \Op {D}\right\rangle^2 )}{4\left\langle \Op {L}\right\rangle ^{2}
+\omega^{2}\left\langle \Op {D}\right\rangle ^{2}-\left(\hbar\omega-2\left\langle \Op {H}\right\rangle \right)^{2}}
~~.
\label{eq:betainver}
\end{equation}
Equations (\ref{eq:gamainver}) and (\ref{eq:betainver}) make explicit the relation between the state of the system 
$\Op \rho $ (by equation (\ref{eq:product})) and the thermodynamical
observables $\langle \Op {H} \rangle$ ,$\langle \Op {D} \rangle$ and $\langle \Op {L} \rangle$.

\subsection{Entropy balance}

In thermodynamics the entropy $S$ is a state variable. 
Shannon introduced entropy as a measure
of the missing information required to define a probability distribution $\bf p$ \cite{shanon48}.
The information entropy can be applied to a complete quantum measurement of an observable 
represented by the operator $\Op O$ with possible outcomes $p_j$:
\begin{equation}
S_{\Op O}=-k_B \sum_j ~p_j \ln p_j~,
\label{eq:obentropy}
\end{equation}
where $p_j= Tr \{ \Op P_j \Op \rho \}$. 
The projections $\Op P_j$ are defined using the spectral decomposition theorem
$\Op O= \sum_j \lambda_j \Op P_j$, where $\lambda_j$ are the eigenvalues of the operator
$\Op O$. $S_{\Op O}$ is then the measure of information gain obtained by the measurement.
The Von Neumann entropy is equivalent to the minimum entropy $S_{\Op Y_n}$ associated with 
a complete measurement of the state $\Op \rho$, by the observable $\Op Y_n$ where the set 
of operators $\Op Y_n$ includes all possible non degenerate operators in the Hilbert space. 
The operator that minimizes the entropy commutes with the state $[\Op \rho, \Op Y_{min}]=0$.
Obviously $ S_{VN} \le S_{\Op O}$. This supplies the interpretation that $S_{VN}$ is the minimum
information required to completely specify the state $\Op \rho$. 

The primary thermodynamic variable for the heat engine is energy.
The entropy associated with the measurement of energy $S_E = S_{\Op H}$ in general
differs from the  Von Neumann entropy $ S_{E} \ge S_{VN}$. 
Only when $\Op \rho$ is diagonal in the energy representation, such as in thermal equilibrium (\ref{eq:req}), $S_{E}= S_{VN}$. 

The Von Neumann entropy is invariant under a unitary evolution \cite{Alhassid}. 
This is the result of the property of unitary transformations 
where the set of eigenvalues of $\Op \rho' =\Op U \Op \rho \Op U^{\dagger}$ 
is equal to the set of eigenvalues of  $\Op \rho$. Since the von Neumann entropy $S_{VN}$ 
is a functional of the eigenvalues of $\Op \rho$ it becomes invariant
to  any unitary transformation.
In particular when applying a unitary transformation  generated by  
the Hamiltonian $\Op H'$ to a thermal state $\Op \rho (0) = \exp(-\Op H /k_b T)/Z$
then if $[\Op H',\Op \rho(0)] ~~\ne ~0$, the unitary transformation
will always increase the energy entropy,  ${\cal S}_E(t) \ge {\cal S}_E(0)$. Nevertheless  
the Von Neumann entropy will stay invariant ${\cal S}_{VN}(t)={\cal S}_{VN}(0)= {\cal S}_E(0)$.

To find the Von Neumann entropy  for the oscillator, 
the product-form of the density operator (\ref{eq:product}) is rewritten as
\begin{equation}
\hat{\rho}=\frac{1}{Z}e^{\chi_{1}{\Op a}^{2} + \chi_{2}{\Op H} + \chi_{1}^*{\Op {a^{\dagger}}}^{2}}~~~,
\label{eq:stateexp}
\end{equation}
where the relation between the coefficients $\chi$ in (\ref{eq:stateexp}) 
and $\gamma$ and $\beta$ in the product form (\ref{eq:product}) 
are evaluated in \ref{sec:relation} , equation (\ref{ea:relation}).

Employing (\ref{eq:stateexp}) the  Von Neumann entropy becomes:
\begin{eqnarray}
{\cal S}_{VN}=-\chi_{2}\left\langle \mathbf{\Op {H}}\right\rangle 
+\mathrm{Im}\left(\chi_{1}\right)\frac{1}{\hbar}\left\langle \mathbf{\Op {D}}\right\rangle 
+\mathrm{Re}\left(\chi_{1}\right)\frac{2}{\hbar\omega}\left\langle \mathbf{\Op{L}}\right\rangle 
+\ln \left(\frac{\mathrm{csch}\left(\frac{\beta\hbar\omega}{2}\right)}{2\sqrt{1-\frac{4\gamma\gamma^{*}}
{\left(e^{\beta\hbar\omega}-1\right)^{2}}}}\right)
\label{eq:vonneumann}
\end{eqnarray}
Equation (\ref{eq:vonneumann}) shows that ${\cal S}_{VN}$ is a functional of
the thermodynamical variables $\langle {\Op {H}}\rangle$, 
$\langle {\Op {D}}\rangle$ and $\langle \Op{L}\rangle$. 

The energy entropy ${\cal S}_E$ of the oscillator (not in equilibrium) 
is found to be equivalent to the entropy of
an oscillator in thermal equilibrium with the same energy expectation:
\begin{equation}
{\cal S}_E~~=~~\frac{1}{\hbar\omega}\left(\left\langle {\Op {H}}\right\rangle 
+\frac{\hbar\omega}{2}\right)
\ln\left(\frac{2\left\langle \mathbf{\hat{H}}\right\rangle + \hbar \omega}
{2\left\langle \mathbf{\hat{H}}\right\rangle -\hbar\omega}\right)
-\ln\left(\frac{2\hbar\omega}{2\left\langle \mathbf{\hat{H}}\right\rangle -\hbar\omega}\right)~~.
\label{eq:enentrop1}
\end{equation}
${\cal S}_E $ in (\ref{eq:enentrop1}) is completely determined by the energy expectation 
$E=\left\langle \mathbf{\Op {H}}\right\rangle$. As an extreme example, for  a squeezed state
${\cal S}_{VN}=0$ and ${\cal S}_E \ge 0$.

In a macroscopic working medium, the internal temperature
can be defined from the entropy and energy variables
$1/T_{int} = \left(\frac{\partial S}{\partial E}\right)_V$ at constant volume.
For the quantum Otto cycle, $S_E$ is used to define the inverse internal temperature
$1/T_{int} = \left(\frac{\partial S_E}{\partial E}\right)_{\omega}$. 
$T_{int}$ is a generalized  temperature appropriate for non  equilibrium density 
operators $\Op \rho$. Using this definition the internal temperature $T_{int}$ of the oscillator 
working medium can be calculated implicitly from the energy expectation:
\begin{equation}
E=\frac{1}{2}\hbar\omega\coth\left(\frac{\hbar\omega}{2 k_B T_{int}}\right)~~,
\end{equation}
which is equivalent to the relation between temperature and energy in thermal equilibrium.

The entropy expressions (\ref{eq:vonneumann}) and (\ref{eq:enentrop1}) for the working medium illustrate that entropy changes along the cycle
trajectory. Once the cycle reaches a limit cycle (Cf. Sec. (\ref{subsec:limit})) then the internal 
entropy changes become cyclic. As a result any entropy change becomes zero \cite{k201}.
The irreversible character of the operation concentrates therefore
in the entropy production in the thermal baths. Assuming a structureless reservoir, the entropy production
is calculated from the heat transfer from the working medium:
\begin{equation}
\Delta S_u ~~=~~ \Delta S_h +\Delta S_c 
~~=~~-~\frac{1}{T_h}\int_0^{\tau_h} \dot {\cal Q}_h dt
~-~\frac{1}{T_c}\int_0^{\tau_c} \dot {\cal Q}_c dt 
\;.
\label{eq:entprod}
\end{equation}
where $T_h$ and $T_c$ are the hot and cold bath temperatures and $\dot {\cal Q}_h$ $\dot {\cal Q}_c$ 
are the heat fluxes Cf. equation (\ref{eq:firstlaw}).

\section{Solving the dynamics of the quantum Otto cycle}
\label{sec:otodynamics}

The strategy to solve the dynamics of the working fluid along
the engine's cycle of operation relies on obtaining closed form
equations for the quantum thermodynamical observables and on using them to reconstruct the state
of the system using the maximum entropy relation (\ref{eq:state}).

\subsection{Heisenberg dynamics on the  Isochores}

The working medium dynamics on the {\em isochore}s represent an approach to thermal equilibrium where the 
Hamiltonian $\Op H$ is constant ($\omega =\omega_{h/c}$ is constant). 
The Heisenberg equations of motion for an operator $\Op X$ are
\begin{eqnarray}
\dot {\Op X} 
=\frac{i}{\hbar}[\Op H,\Op X]+
k_{\downarrow}(\Op a^{\dagger}\Op X \Op {a}-\frac{1}{2}\{\Op {a}^{\dagger}\Op {a},\Op X\})+
k_{\uparrow}(\Op{a}\Op X \Op {a}^{\dagger}-\frac{1}{2}\{\Op{a}\Op{a}^{\dagger},\Op X\})
\label{eq:hieniso}
\end{eqnarray}
Equation (\ref{eq:hieniso}) is the analogue of (\ref{eq:lvn2}) and (\ref{eq:dissipative}) 
in the Schr\"odinger frame.

For the dynamical set of observables, the equations of motion become
\begin{eqnarray}
\frac{d}{dt}\left(\begin{array}{c}
\Op H\\
\Op L\\
\Op D\\
\Op I
\end{array}\right)(t)=
\left(\begin{array}{cccc}
-\Gamma & 0 & 0 & \Gamma \langle \Op H \rangle_{eq}\\
0 & -\Gamma & -\omega^{2} & 0 \\
0 & 4 & -\Gamma & 0\\
0&0&0&0
\end{array}\right)
\left(\begin{array}{c}
\Op H\\
\Op L\\
\Op D\\
\Op I
\end{array}\right)(t)
\label{eq:motion}
\end{eqnarray}
where $\Gamma=k_{\downarrow}-k_{\uparrow}$ is the heat conductance and 
$k_{\uparrow}/k_{\downarrow}= e^{-\hbar\omega/k_B T}$ obeys detailed balance where
$\omega=\omega_{h/c}$ and $T=T_{h/c}$ 
are  defined for the hot or cold bath respectively. From (\ref{eq:firstlaw}) the heat current can be identified as:
\begin{eqnarray} 
\dot {\cal Q}=-\Gamma ( \langle \Op H \rangle -\langle \Op H \rangle_{eq})~~=~~\Gamma \frac{\hbar \omega}{2}
\left( \coth ( \frac{\hbar \omega}{2 k_B T_B})-\coth ( \frac{\hbar \omega}{2 k_B T_{int}}) \right)~~.
\label{eq:hettrans}
\end{eqnarray}
where $T_B$ is the bath temperature.
In the high temperature limit the heat transport law becomes Newtonian: 
$ \dot {\cal Q}=\Gamma (T_B -T_{int})$.

The solution of the {\em isochore} dynamics (\ref{eq:motion}) becomes:
\begin{eqnarray}
\Op H(t) & = & e^{-\Gamma t}(\Op H(0)-\Op I \langle \Op H \rangle_{eq})+\Op I \langle \Op H \rangle_{eq}
\label{eq:isodynamics}
\end{eqnarray}
\begin{eqnarray}
\left(\begin{array}{c}
\Op L\\
\Op D\end{array}\right)(t) 
& = & e^{-\Gamma t}\left(\begin{array}{cc}
\cos\left(2\omega t\right) & -\frac{\omega}{2}\sin\left(2\omega t\right)\\
\frac{2}{\omega}\sin \left( 2\omega t\right) & \cos\left(2\omega t\right)\end{array}\right)\left(\begin{array}{c}
\Op L\\
\Op D\end{array}\right)(0)
\end{eqnarray}
where the equilibrium value of the energy is 
$E= \langle \Op H \rangle_{eq}=\frac{\hbar\omega}{2}\coth(\frac{\hbar\omega}{2 k_B T})$. 

The equilibration dynamics are characterized by the energy exponentially relaxing to its equilibrium value. 
The Lagrangian $\Op L$ and the momentum-position correlation $\Op D$ show a damped oscillation with 
a frequency of $2\omega$ to an equilibrium value of zero. The identity operator $\Op I$ becomes a constant of motion representing the conservation of norm.

\subsection{The dynamics on the  Adiabats}

On the {\em adiabat}s the oscillator frequency changes from $\omega_h$ to $\omega_c$ on the power
expansion branch and from $\omega_c$ to $\omega_h$ on the compression branch. As a result
the Hamiltonian is explicitly time dependent and $[\Op H (t),\Op H(t')] \neq 0$.
Under these conditions the Heisenberg equations of motion (\ref{eq:heisenberg}) for the dynamical set of operators become
\begin{equation}
\frac{d}{dt}\left(\begin{array}{c}
\Op H\\
\Op L\\
\Op D\\
\Op I
\end{array}\right)(t)=
\left(\begin{array}{cccc}
\frac{\dot{\omega}}{\omega} & -\frac{\dot{\omega}}{\omega} & 0&0\\
-\frac{\dot{\omega}}{\omega} & \frac{\dot{\omega}}{\omega} & -\omega^{2}&0\\
0 & 4 & 0&0\\
0&0&0&0
\end{array}\right)
\left(\begin{array}{c}
\Op H\\
\Op L\\
\Op D\\
\Op I
\end{array}\right)(t)~.
\label{eq:adiabatdy}
\end{equation}
In general all operators in (\ref{eq:adiabatdy}) are dynamically coupled. The coupling
can be characterized by the nonadiabatic parameter $\alpha=\frac{\dot{\omega}}{\omega}$.
It is important to notice that $\alpha$ has no explicit dependence on the mass of the oscillator.
This means that the nonadiabatic character is universal. For example from
the first law (\ref{eq:firstlaw}) the power can be identified as
\begin{equation}
{\cal P} = \alpha \left( \langle \Op H \rangle -\langle  \Op L \rangle \right)
~.
\label{eq:pl}
\end{equation}
The power on the {\em adiabats} (\ref{eq:pl}) can be decomposed to the "useful" external power
${\cal P}_{ex}= \alpha \langle  \Op H \rangle $ and to the power invested to counter friction
${\cal P}_{f}= -\alpha \langle  \Op L \rangle $. 
It is ``friction'' in the sense that it reduces the power and is zero 
when the motion is infinitely slow and increases with speed (see below). 
The translation of this effect into the actual reduction of power, 
however, cannot be separated from the dissipative branches.

A solution of the dynamics depends on an explicit dependence of $\omega(t)$ on time, 
and it can always be carried out numerically.
An illustrative example is found when the non-adiabatic parameter $\alpha=\frac{\dot \omega}{\omega}$ 
is constant during the adiabatic branches. This leads to the explicit time dependence 
$\omega(t)=\omega_0 e^{\alpha t}=\omega_h \left( \frac{\omega_c}{\omega_h} \right)^{t/\tau_{hc}}$ on the power {\em adiabat}, and a similar expression for the compression {\em adiabat}.
Under these conditions the time derivative in (\ref{eq:adiabatdy}) becomes stationary 
allowing a closed 
form solution to be obtained by diagonalizing the matrix in (\ref{eq:adiabatdy}). 

Two simplifying limits exist. The first is the quasistatic limit, which is defined by  $\alpha \ll \omega$.
In this case the fast motion (neglecting $\alpha$) can be evaluated first in (\ref{eq:adiabatdy}) leading to
a zeroth order expression for $\Op L$ and $\Op D$. From that expression one can construct an ansatz correct to first order in $\frac{\alpha}{\omega}$:
\begin{eqnarray}
\begin{array}{lcl}
\Op L(t) & = & \frac{2\omega(t)}{\omega_0}\Op L(0)\cos(\phi)
-\omega(t)\left(\frac{\alpha}{2\omega_0^2}\Op H(0)+\frac{1}{2}\Op D(0)\right)\sin(\phi)\\

\Op D(t) & = & -\frac{\alpha}{\omega(t)\omega_0}\Op H(0)
+\left(\frac{\alpha}{\omega_0^2}\Op H(0)+ \Op D(0)\right)\cos(\phi)
+\frac{2}{\omega_0} \Op L(0) \sin(\phi)~~~,
\end{array}
\end{eqnarray}
where $\phi(t)=\int_0^t 2 \omega(t)dt=\frac{2(\omega(t)-\omega_0)}{\alpha}$ 
is the accumulated phase. 
For small $\frac{\alpha}{\omega}$, $\phi$ diverges leading to
fast oscillation of $\Op L(t)$. As a result $\Op L$ can be averaged out in the
equation of the slow variable $\Op H$ leading to:
\begin{equation}
\frac{d \Op H}{dt}= \alpha \Op H
\label{eq:adiabat3}
\end{equation}
with the solution $\Op H(t) = \frac{\omega(t)}{\omega_h}\Op H(\omega_0=\omega_h)$ 
for the power {\em adiabat} and $\Op H(t) = \frac{\omega(t)}{\omega_c}\Op H(\omega_0=\omega_c)$ 
for the compression {\em adiabat}. 

The outcome of the quasistatic conditions is that 
the Hamiltonian follows adiabatically 
the change in frequency $\omega$. This limit is equivalent to the invariance of the number
operator $\Op N = \frac{1}{\hbar \omega}\Op H - \frac{1}{2}\Op I$ in the evolution, i.e. 
$\langle \Op N \rangle = constant$. In this limit the power invested to counter friction
is zero, ${\cal P}_{f}=0$.

A higher order expression for the energy $\Op H$ can be obtained by integration, and will be useful in the consideration of friction. To second order in $\frac{\alpha}{\omega}$ the energy $\Op H$  on the {\em adiabat} becomes
\begin{eqnarray}
\nonumber
8 \omega_0^3 \omega(t) \Op H(t) & = & \Op H (0) \left(-\alpha^3 \omega_0
\sin(\phi)-2\alpha^2\omega_0\omega(t)\cos(\phi)+2\alpha^2 \omega(t)^2+8\omega_0^2\omega(t)^2 \right)\\
\nonumber
& + & \Op L(0)
\left(-4\alpha\omega_0^2\omega(t)\sin(\phi)+2\alpha^2\omega_0^2\cos(\phi)-2\alpha^2\omega(t)^2 \right)
\\ & + & \Op D(0)
\left(-\alpha^2\omega_0^3\sin(\phi)-2\alpha\omega_0^3\omega(t)\cos(\phi)+2\alpha\omega_0^2\omega(t)^2\right)~~.
\end{eqnarray}
Averaging over the fast oscillation, $\bar{  \sin(\phi)} =0$ leads to:
\begin{eqnarray}
\Op H (t) & \approx & \Op H(0)
\frac{\omega(t)}{\omega_0}\left (1~+\left(\frac{\alpha}{2\omega_0}\right)^2~\right)-\Op L (0)
\frac{\omega(t)}{ \omega_0}\left( \frac{\alpha}{2 \omega_0} \right)^2+ \Op D(0)
\alpha\frac{\omega(t)}{4 \omega_0}~.
\end{eqnarray}

The other extreme condition is the sudden limit when $\alpha \gg \omega$.
The fast variables can be integrated first leading to
\begin{equation}
\left(\begin{array}{c}
\Op H\\
\Op L\\
\end{array}
\right)(t)=
\frac{1}{2}\left(\begin{array}{cc}
1+e^{2 \alpha t} & 1-e^{2 \alpha t} \\
1-e^{2 \alpha t} & 1+e^{2 \alpha t} \\
\end{array}\right)
\left(\begin{array}{c}
\Op H\\
\Op L\\
\end{array}\right)(0)
\label{eq:sudden}
\end{equation}
and $\Op D$ staying constant up to first order in $\frac{\omega}{\alpha}$.
The dynamics can be translated to the final expectation at the end of the {\em adiabat}:
\begin{equation}
\left(\begin{array}{c}
\Op H\\
\Op L\\
\end{array}
\right)_{final}=\frac{1}{2}\left(\begin{array}{cc}
1+(\frac{\omega_{f}}{\omega_{i}})^2 & 1-(\frac{\omega_{f}}{\omega_{i}})^2 \\
1-(\frac{\omega_{f}}{\omega_{i}})^2 & 1+(\frac{\omega_{f}}{\omega_{i}})^2 \\
\end{array}\right)\left(\begin{array}{c}
\Op H\\
\Op L\\
\end{array}\right)_{initial}~~~,
\label{eq:sudden1}
\end{equation}
where $\omega_{i/f}$ is either $\omega_h$ or $\omega_c$ depending on whether
the evolution takes place on the power or on the compression {\em adiabat}.

The dynamics of the working fluid on the different branches of the engine
can be summarized by a branch propagator ${\cal U}_k$ which maps the set
of operators $\vec {\Op A}$ across the branch. An example is the hot {\em isochore}
$\vec {\Op A}(\tau_h) = {\cal U}_h \vec {\Op A}(0)$.

\section{The Engine In Action}
\label{sec:engineaction}

\begin{figure}
\vspace{0.5cm}
\center{\includegraphics[width=0.65\textwidth]{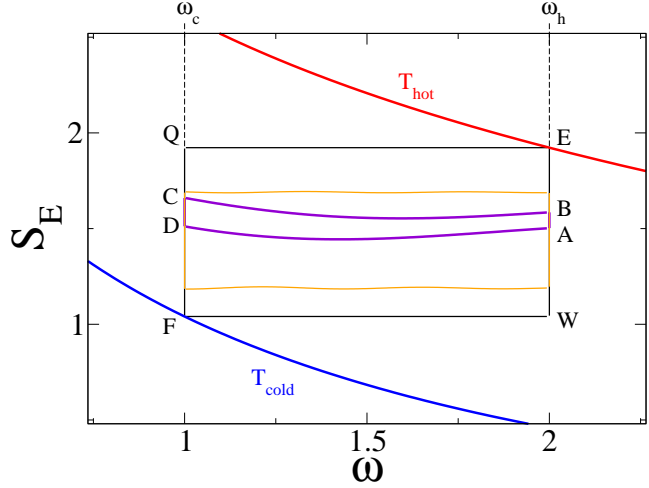}}

\caption{\label{fig:1} The limit cycle of the engine in the (${\cal S}_E$ ,$\omega$) 
plane. The finite time cycle is indicated by the points {\bf A B C D}.
The cycle with infinite time allocation to all branches is indicated
by the points {\bf W E Q F}. Another long finite time cycle is indicated in yellow. 
The equilibrium isotherms of the hot bath intersecting at point {\bf E}
and the cold bath intersecting at point {\bf F}, are also shown.
$\hbar \omega_h=2$, $\hbar \omega_c=1$, $k_B T_h=5$, $k_B T_c=1$, 
$\Gamma =0.03$ in units of $\omega_c$. $\tau_h=6~,~\tau_c=12~,~\tau_{hc}=\tau_{hc}=1$
in units of $\omega_c^{-1}$.}
\end{figure}

We can now turn to look at the engine cycle as a whole. The four branches of the engine can be summarized as follows:
\begin{enumerate}
\item{The hot {\em isochore}: $A \rightarrow B$ in figure \ref{fig:1}, characterized by 
$\omega_h$ the working medium frequency,  the temperature $T_h$
of the hot bath, $\Gamma_h$ the  heat transfer rate, 
$\langle \Op H \rangle_{eq}(T_h)$ the equilibrium energy of the working fluid point E. 
The external control is the time allocation $\tau_h$ which determines the extent of equilibration.
The dynamics of the operator set is represented by the propagator ${\cal U}_h$.}
\item{The power {\em adiabat}: $B \rightarrow C$ in figure \ref{fig:1}, characterized by the change 
from $\omega_h$ to $\omega_c$ of the working medium frequency. The external control is
the time allocation $\tau_{hc}$ translated in to the adiabatic parameter $\alpha_{hc}$.
The dynamics of the operator set is represented by the propagator ${\cal U}_{hc}$.}
\item{The cold {\em isochore}: $C \rightarrow D$ in figure. \ref{fig:1}, characterized by the 
$\omega_c$ the working medium frequency,  the temperature $T_c$
of the hot bath, $\Gamma_c$ the  heat transfer rate, 
$\langle \Op H \rangle_{eq}(T_c)$ the equilibrium energy of the working fluid point F. 
The external control is the time allocation $\tau_c$ which determines the extent of equilibration.
The dynamics of the operator set is represented by the propagator ${\cal U}_c$.}
\item{The compression {\em adiabat}: $D \rightarrow A$ in figure \ref{fig:1}, characterized by the change 
from $\omega_c$ to $\omega_h$ of the working medium frequency. The external control is
the time allocation $\tau_{ch}$ translated to the adiabatic parameter $\alpha_{ch}$.
The dynamics of the operator set is represented by the propagator ${\cal U}_{ch}$.}
\end{enumerate}

\subsection{The limit cycle}
\label{subsec:limit}

Once the engine is ignited it is expected that it
will settle down to a smooth mode of operation determined by the control parameters. 
This asymptotic cycle will be termed the {\bf limit cycle}. 

The dynamics of the four strokes of the cycle can be represented by
the cycle propagator:
\begin{equation}
{\cal U}_{cyc}~~=~~ {\cal U}_{ch}~{\cal U}_{c}~{\cal U}_{hc}~{\cal U}_{h}~~~.
\label{eq:cycprop}
\end{equation}
It is therefore expected that each point of the limit cycle trajectory is
invariant to the propagator 
${\cal U}_{cyc} \vec {\Op A}_{limit} =\vec {\Op A}_{limit}$ (we choose point $A$).
The cycle propagator is a product of the branch propagators. They belong
to the class of completely positive maps, ${\cal U}_{cyc}$ therefore, 
is also a completely positive map. It has been proven that if $\vec {\Op A}_{limit}$
is unique then a repeated application of ${\cal U}_{cyc}$ will lead 
monotonically to the limit cycle \cite{k116,k201}. This means
that the limit cycle has to close. Consequently all variables are cyclic.
This property allows to study the optimal performance of the operating engine.

\begin{figure}
\center{\includegraphics[width=0.65\textwidth]{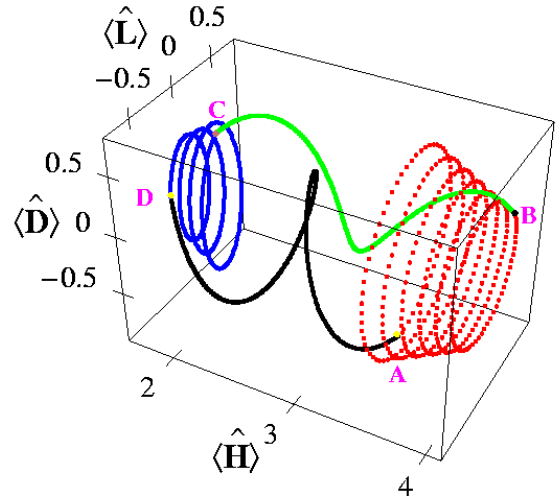}}

\vspace{-2.cm}
\caption{\label{fig:2} A limit cycle trajectory in the $\langle \Op H \rangle$,
$\langle \Op L \rangle$ and $\langle \Op D \rangle$ coordinates. $\bf A \rightarrow \bf B$
is the hot {\em isochore}, $\bf B \rightarrow \bf C$ is the power {\em adiabat},
$\bf C \rightarrow \bf D$ is the cold {\em isochore} and $\bf D \rightarrow \bf A$
is the expansion {\em adiabat}. }
\end{figure}

A typical limit cycle trajectory is shown in figure \ref{fig:2}. 
The oscillation of $\langle \Op L \rangle$ and $\langle \Op D \rangle$ on the {\em isochores}
with a different frequency $\omega_h$ or $\omega_c$ is clearly seen. The canonical invariance condition
is preserved by the limit cycle. Moreover even if the initial state of the system does not
belong to the class of canonical invariant states after a few cycles it will approach
monotonically such a state \cite{k201}. 

\subsection{Optimizing the operation of the engine}

Finite time thermodynamics allows us to optimize power as well as efficiency. We will assume, for simplicity, that the external parameters are fixed. These include the bath parameters $T_h$ and $T_c$ as well as
the heat transport rates $\Gamma_h$ and $\Gamma_c$. At first
the frequencies $\omega_h$ and $\omega_c$ and with them the
compression ratio ${\cal C} =\omega_h/\omega_c$ are also set.
The optimization is carried out with respect to the time allocations
on each of the engine's branches:
$\tau_h$, $\tau_{hc}$, $\tau_c$, and $\tau_{ch}$. This sets the total cycle time
$\tau = \tau_h+\tau_{hc}+\tau_c+\tau_{ch}$.

The maximum efficiency limit is obtained for infinite time allocations on all branches.
This  cycle maximizes the work output. Its trajectory passes through the points {\bf W E Q F} 
in figure \ref{fig:1}, and the efficiency reaches the value $\eta_o= 1 - \frac{\omega_c}{\omega_h}$. 
Since for this limit the cycle time $\tau$ is infinite, obviously the power of this cycle is zero.

Optimizing the power is far more interesting. It requires a finite cycle time $\tau$. As a result the optimizer 
has to compromise and shorten the time allocated to equilibration on the {\em isochores}. 
In addition restricting the time on the {\em adiabat}s results in nonadiabatic conditions that expose the nature of friction.

\subsubsection{Optimizing the operation of the engine for quasistatic conditions.}

The power optimization in the quasistatic regime is determined by
heat transport from the hot and cold baths.
To reach quasistatic conditions, ample time is allocated to the {\em adiabats}  
(Cf. equation (\ref{eq:adiabat3})). As a consequence,
the Number operator becomes invariant on the {\em adiabats}. 

The time allocations on the
{\em isochores} determine the change in the number operator 
$N=\langle\Op N \rangle$ (Cf. equation (\ref{eq:isodynamics})):
$N^{\bf B} = e^{-\Gamma_h \tau_h} \left( N^{\bf A} - N_{eq}^h \right) +N_{eq}^h $ 
on the hot {\em isochore}, 
where $N^{\bf B}$ is the number expectation value at the end of the hot {\em isochore}, 
$N^{\bf A}$ at the beginning and $N_{eq}^h$ is the equilibrium value point {\bf E}. 
A similar expression exists for the cold {\em isochore}. 

The work in the limit cycle becomes
\begin{eqnarray}
{\cal W}_q ~~= ~~ E^{\bf C}-E^{\bf B}+E^{\bf A}-E^{\bf D}
~~  = ~~ \hbar(\omega_{c}-\omega_{h})(N^{\bf B}-N^{\bf D})
\label{eq:11}
\end{eqnarray}
where the convention of the sign of the work for a working engine is negative, in correspondence with Callen \cite{callen}.

The heat transport from the hot bath becomes
\begin{eqnarray}
{\cal Q}_h ~~= ~~ E^{\bf B}-E^{\bf D}
~~  = ~~ \hbar\omega_{h}(N^{\bf B}-N^{\bf D})~.
\label{eq:heat11}
\end{eqnarray}

The efficiency $\eta$ becomes independent of
the time allocations:
\begin{equation}
\eta_q ~~=~~\eta_o~~=~~ 1- \frac{\omega_c}{\omega_h} ~~=~~ 1 - \frac{1}{\cal C}~.
\label{eq:effic}
\end{equation}

In the limit cycle for adiabatic conditions $N^{\bf D}=N^{\bf A}$
which leads to the relation
\begin{equation}
N^{\bf B}=\frac{(e^{\Gamma_{c}\tau_{c}}-1)}{1- e^{\Gamma_{c}\tau_{c}
+\Gamma_{h}\tau_{h}}}(N_{eq}^{h}-N_{eq}^{c})+N_{eq}^{h}~.
\label{eq:number1}
\end{equation}
By demanding a closed cycle the number change $N^{\bf B}-N^{\bf D}$ is calculated,
leading to:
\begin{eqnarray}
\label{eq:quasistatic work}
{\cal W}_q & = & \hbar(\omega_{h}-\omega_{c})(N_{eq}^{h}-N_{eq}^{c})\frac{(e^{x_{c}}-1)(e^{x_{h}}-1)}
{1-e^{x_{c}+x_{h}}}\label{eq:quasistatic-work}\\
 & \equiv & -G_W(T_{c},\omega_{c},T_{h},\omega_{h})F(x_{c},x_{h})~~,
\nonumber
\end{eqnarray}
where the scaled time allocations are defined $x_{c}\equiv\Gamma_{c}\tau_{c}$, and $x_{h}\equiv\Gamma_{h}\tau_{h}$.
The work ${\cal W}_q$ (\ref{eq:quasistatic-work}) becomes a product 
of two functions: $G_W$ which is a function of the static constraints of the engine
and $F$ which describes the heat transport on the {\em isochores}. Explicitly the function $G_W$ is
\begin{equation}
G_W(T_{c},\omega_{c},T_{h},\omega_{h})=\frac{\hbar}{2}\left(\omega_{h}-\omega_{c}\right )
\left( \coth\left(\frac{\hbar\omega_h}{2 k_B T_h}\right)-\coth\left(\frac{\hbar\omega_c}{2 k_B T_c}\right)\right)~~.
\label{eq:gfunct}
\end{equation} 

$ 0 \le F \le 1$, therefore
for the engine to produce work $G_W \ge 0$. 
The first term in (\ref{eq:gfunct}) is positive. Therefore $G_W \ge 0$ translates in
to the restriction that  $\frac{\omega_c}{\omega_h} \ge \frac{T_c}{T_h}$ 
or in terms of the compression ratio $ 1 \le {\cal C} \le \frac{T_h}{T_c}$.
This is equivalent to the  statement that the maximum efficiency of the Otto cycle 
is smaller than the Carnot efficiency $\eta_{o} \le \eta_{c}$. 

In the high temperature limit when
$\frac{ \hbar \omega}{ K_bT} \ll 1$, $G_W$ simplifies to
\begin{equation}
G_W ~~=~~ k_B T_c(1-{\cal C})~+~k_BT_h(1-{\cal C}^{-1})~. 
\label{eq:hightempG}
\end{equation}
In this case the work ${\cal W}_q = -G_W ~F$ can be optimized 
with respect to the compression ratio ${\cal C} = \frac{\omega_h }{\omega_c}$  
for  fixed bath temperatures. The optimum is found at ${\cal C}=\sqrt{\frac{T_h}{T_c}}$. 
As a  result the efficiency at maximum power for high temperatures becomes
\begin{equation} 
\eta_q = 1-\sqrt{\frac{T_c}{T_h}}~,
\label{eq:endoeffic}
\end{equation}
which is  the well known efficiency at maximum power of an endoreversible engine 
\cite{curzon75,k85,k24,jeff89,broeck05}. Note that these results indicate greater validity 
to the Curzon-Ahlborn result from what their original derivation \cite{curzon75} indicates.

The function $F$ defined in (\ref{eq:quasistatic work}) characterizes the heat transport to the working
medium. As expected $F$ maximizes when infinite time is allocated to the {\em isochores}.
The optimal partitioning of the time allocation between the hot and cold {\em isochores}
is obtained when:
\begin{equation}
\Gamma_{h}(\cosh(\Gamma_{c}\tau_{c}) -1 )=\Gamma_{c}(\cosh(\Gamma_{h}\tau_{h})-1)~~.
\label{eq: d-tau optimum work}
\end{equation}
If (and only if) $\Gamma_{h}=\Gamma_{c}$
the optimal time allocations on the {\em isochores} becomes  $\tau_{h}=\tau_{c}$. 

Optimizing the total cycle power output ${\cal P}$ is equivalent to optimizing $F/\tau$ since
$G_W$ is determined by the engine's external constraints. 
The total time allocation $\tau=\tau_{iso}+\tau_{adi}$ is partitioned to the time on the
{\em adiabat}s $\tau_{adi}$ which is limited by the adiabatic condition,
and the time $\tau_{iso}$ allocated to the {\em isochores}. 

Optimizing the time allocation on the {\em isochore}s subject to (\ref{eq: d-tau optimum work}) leads to the optimal condition
\begin{equation}
\Gamma_{c}\tau(\cosh(\Gamma_{h}\tau_{h})-1)=
\sinh(\Gamma_{h}\tau_{h}+\Gamma_{c}\tau_{c})-\sinh(\Gamma_{c}\tau_{c})-\sinh(\Gamma_{h}\tau_{h})~.
\label{eq:quasistatic optimal tau}
\end{equation}
When $\Gamma_{h}=\Gamma_{c}\equiv\Gamma$ this expression simplifies to:
\begin{equation}
2x+\Gamma\tau_{adi}=2\sinh(x)
\label{eq:ftf}
\end{equation}
(where $x=\Gamma_{c}\tau_{c}=\Gamma_{h}\tau_{h}$). For small $x$ equation (\ref{eq:ftf})
can be solved leading to the optimal time allocation on the {\em isochores}: 
$\tau_c=\tau_h \approx \left(\Gamma \tau_{adi}/3\right)^{\frac{1}{3}}/\Gamma$. Taking into consideration
the restriction on the adiabatic condition this time can be estimated to be:
$\tau_c=\tau_h \approx \frac{1}{\Gamma} \left(\frac{\Gamma}{\omega} \right)^{\frac{1}{3}}$.
When the heat transport rate $\Gamma$ is sufficiently large, the optimal power
conditions lead to the Bang-Bang solution where vanishingly small time is allocated
to all branches of the engine \cite{k116}.

For the conditions $\Gamma\equiv\Gamma_c=\Gamma_h$, and restricting the time allocation on the 
{\em adiabat} to be 
$\tau_{adi}=\frac{1}{\omega_c}+\frac{1}{\omega_h}$,  the quasistatic optimal power production 
as a function of cycle time $\tau$ becomes:
\begin{eqnarray}
\begin{array}{l}
{\cal P}_q = \hbar (\omega_{h}-\omega_{c} )
\left( \coth\left(\frac{\hbar\omega_h}{2 k_B T_h}\right)-\coth\left(\frac{\hbar\omega_c}{2 k_B T_c}\right)\right)
\frac{\tanh\left(\frac{\Gamma}{4}(\tau-\frac{(\omega_h+\omega_c)}{\omega_h\omega_c})\right)}{2 \tau}
\label{eq:quasistaticpower}
\end{array}
\end{eqnarray}
which is an upper limit to the power production Cf. figures \ref{fig:powerallA}, \ref{fig:powerallB} and \ref{fig:powerallC}.

The entropy production $\Delta S_U$ (\ref{eq:entprod}) reflects the irreversible character of the engine.
In quasistatic conditions the irreversibility is completely associated with the
heat transport. $\Delta S_U$ can also be factorized to a product of two functions:
\begin{equation}
\Delta S_u ~~=~~G_S (T_{c},\omega_{c},T_{h},\omega_{h}) F(x_{c},x_{h})~~,
\label{eq:entroporod}
\end{equation}
where $F$ is identical to the $F$ function defined in (\ref{eq:quasistatic-work}).
The function $G_S$ becomes:
\begin{equation}
G_S (T_{c},\omega_{c},T_{h},\omega_{h}) ~~=~~\frac{1}{2}
\left(\frac{\hbar \omega_h}{k_B T_h}-\frac{\hbar \omega_c}{k_B T_c} \right)
\left(\coth\left(\frac{\hbar\omega_c}{2 k_B T_c}\right) -\coth\left(\frac{\hbar\omega_h}{2 k_B T_h}\right)\right)~~.
\label{eq:entroporod2}
\end{equation}
Due to the common $F(x_{c},x_{h})$ function the entropy production has the same dependence 
on the time allocations $\tau_h$ and $\tau_c$ as the work ${\cal W}$ \cite{k152}.
As a consequence maximizing the power will also maximize the entropy production 
rate $\Delta S_U/\tau$. Note that entropy production is {\em always} positive, even for cycles that produce no work as their compression ratio $\cal C$ is too large, which is a statement of the second law.

The dependence of the $G_s$ function on the compression ratio can be simplified
in the high temperature limit leading to:
\begin{equation}
G_S ~~=~~ {\cal C} \frac{T_c}{T_h}+{\cal C}^{-1}\frac{T_h}{T_c}-2~~~,
\label{eq:entroporod3}
\end{equation}
which is a monotonic decreasing function in the range $ 1 \le {\cal C} \le \frac{T_h}{T_c}$ 
that reaches a minimum at the Carnot boundary when ${\cal C}=\frac{T_h}{T_c}$.
\begin{figure}
\vspace{0.8cm}
\center{\includegraphics[width=0.65\textwidth]{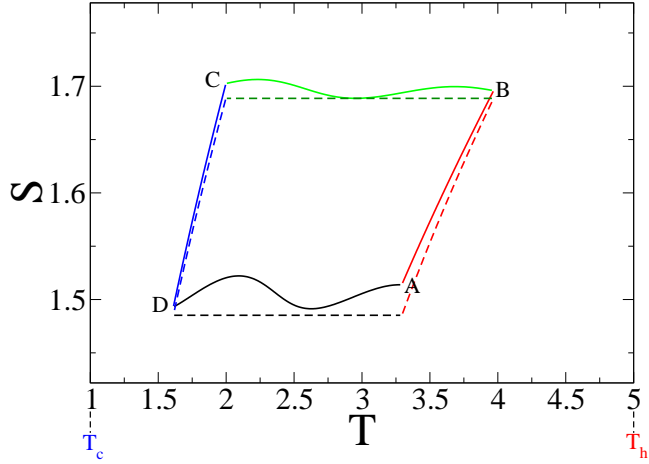}}

\caption{\label{fig:4} The cycle of the engine in the (${\cal S}_E$ ,$T$)
plane (solid lines) and the (${\cal S}_{VN}$ ,$T$) plane (dashed lines). 
The finite time cycle is indicated by the points {\bf A B C D}.
The difference between ${\cal S}_E$ and ${\cal S}_{VN}$ 
decreases at the end of the {\em isochores}. 
The parameters used were: $\hbar \omega_h=2$, $k_B T_h=5$ $\hbar \omega_c=1$, $k_B T_c=1$,
$\Gamma = 0.05$, $\tau_h=~\tau_c=10~,~\tau_{hc}=\tau_{hc}=3$. 
Notice that ${\cal S}_E \ge {\cal S}_{VN}$ and the non monotonic change in ${\cal S}_E$
on the {\em adiabats}.}
\end{figure}

\subsubsection{Lost work required to counter the friction}
\label{subsubsec:fric}
Quasistatic conditions only obtain for infinitely long {\em adiabat}s. To determine the validity and applicability of this limit, we must consider shorter {\em adiabats} and verify our results to higher orders. We must consider finite adiabatic branches which will lead to friction (Cf. equation (\ref{eq:pl})). 
 
The accumulated work required to overcome the friction is defined as:
\begin{equation}
{\cal W}_{f}(t) = -\int_0^{t} \alpha \langle \Op L \rangle dt~~,
\label{eq:wfric}
\end{equation}
for each {\em adiabatic} branch. To second order in the nonadiabatic coupling $\frac{\alpha}{\omega}$,
the accumulated work becomes:
\begin{eqnarray}
{\cal W}_{f}(t) & = &-
\alpha\left(\frac{1}{2\omega_0} \langle \Op L\rangle(0) \sin(\phi)- \left (\frac{1}{2\omega_{0}^2}\alpha
\langle \Op H \rangle(0) +\frac{1}{2} \langle \Op D \rangle(0) \right)\sin^2(\frac{\phi}{2})~\right)~~.
\end{eqnarray}
Averaging on the fast oscillating phase $\phi$ leads to:
\begin{eqnarray}
\bar{{\cal W}_{f}}  & \approx & \frac{\alpha^2}{4\omega_{0}^2}
\langle \Op H \rangle (0) +\frac{\alpha}{4} \langle \Op D \rangle (0)
\end{eqnarray}
When 
$\langle \Op D \rangle (0)=0$ and $\Gamma \rightarrow \infty$, the accumulated work to counter friction 
on both {\em adiabats} for the limit cycle becomes
\begin{eqnarray}
{\cal W}_f ~~ = 
\frac{1}{8}\left( \left(\frac{\alpha} {\omega_h}\right)^2
\hbar \omega_c \mathrm{coth}\left(\frac{\hbar \omega_h}{2 k_B T_h }\right)
+\left(\frac{\alpha}{\omega_c}\right)^2 \hbar \omega_h\mathrm{coth}\left(\frac{\hbar \omega_c}{2 k_B T_c}\right)\right)
~~,
\label{eq:wfric2}
\end{eqnarray}
which  at the high temperature limit simplifies to
\begin{eqnarray}
{\cal W}_f ~~ =~~ 
\frac{k_B T_h}{4}\left(\frac{\alpha}{\omega_c}\right)^2\left(~\mathcal{C}^{-3}~
+~\mathcal{C}\frac{T_c}{T_h}~\right)~~.
\label{eq:hightwfric}
\end{eqnarray}
Equations (\ref{eq:wfric2}) and (\ref{eq:hightwfric}) reflect the  
finite time allocation on the  {\em adiabat} (i.e. the finiteness of $\alpha$) which
always {\em reduces} the produced work relative to the quasistatic case \cite{k176}. 
Remember that  for an engine producing work ${\cal W}={\cal W}_q+{\cal W}_f$ 
should be negative.  
This justifies the use of the term ``friction''. When the friction is sufficiently large, no work can be produced at all in the given engine. This can be seen in figure \ref{fig:work}, showing the quasistatic work ${\cal W}_q$ corrected by ${\cal W}_f$  as a function
of the temperature ratio $r=\frac{T_h}{T_c}$.

The entropy production due to friction is
\begin{eqnarray}
\Delta {\cal S}_f ~~ =~~ \frac{1}{8}\left(\left(\frac{\alpha}{\omega_c}\right)^2 \frac{\hbar\omega_h}{k_BT_h}
\mathrm{coth}\left(\frac{\hbar\omega_c}{2k_B T_c}\right)+\left(\frac{\alpha}{\omega_h}\right)^2 \frac{\hbar\omega_c}{k_BT_c}\mathrm{coth}\left(\frac{\hbar\omega_h}{2k_B T_h}\right)\right)~~.
\label{eq:entfric}
\end{eqnarray}
At the limit of high temperature (\ref{eq:entfric}) simplifies to
$\Delta {\cal S}_f =
\frac{1}{4}\left(\frac{\alpha}{\omega_c}\right)^2 \left({\cal C}\frac{T_c}{T_h}+{\cal C}^{-3}\frac{T_h}{T_c} \right) $.
The nonadiabatic nature of the motion leads to an increase in entropy production in the baths, 
in accordance with our conceptions of ``friction''.

\subsubsection{Optimizing the engine in the sudden limit}
\label{subsubsec:suden}

The extreme case of the performance of an engine with zero time allocation on the {\em adiabat}s 
is dominated by the {\em frictional} terms. These terms arise from the inability
of the working medium to follow adiabatically the external change in potential.
A closed form expression for the sudden limit can be derived based on the 
{\em adiabatic} branch propagator ${\cal U}_{hc}$ and ${\cal U}_{ch}$ (\ref{eq:sudden1}). The expressions are complex and therefore will not be presented.

Insight on the sudden limit can be obtained 
when the heat conductance terms $\Gamma$ become infinite.
The performance therefore is solely determined by the {\em friction}. 
For this limiting case the work per cycle becomes:
\begin{equation}
{\cal W}_s ~~=~~ \frac{(\omega_c -\omega_h)(\omega_c +\omega_h)}{4 \omega_c\omega_h}
 \left(\hbar \omega_c \coth(\frac{\hbar \omega_h}{k_B T_h})
-\hbar \omega_h \coth(\frac{\hbar \omega_c}{k_B T_c})\right)~~.
\label{eq:sudwork1}
\end{equation}
The maximum produced work $-{\cal W}_s$ can be optimized with respect to the compression ratio $\cal C$.
At the high temperature limit: 
\begin{equation}
{\cal W}_s = \frac{1}{2}k_B T_h ( {\cal C}^2-1)(\frac{T_c}{T_h}-\frac{1}{{\cal C}^2})~~.
\label{eq:sudworkh}
\end{equation}
For the quasistatic optimal compression ratio ${\cal C}= \sqrt{ \frac{T_h}{T_c}}$, ${\cal W}_s$ is zero.
The optimal compression ratio for the sudden limit is 
${\cal C}= \left( \frac{T_h}{T_c} \right)^{1/4}$, leading to
the maximal work in the high temperature limit
\begin{equation}
{\cal W}_s ~~=~~ -k_B T_h \left(~ 1-\sqrt{\frac{T_c}{T_h}}~ \right)^2~~~.
\label{eq:sudwork}
\end{equation}
The efficiency at  the maximal work point is
\begin{equation}
\eta_s~~=~~ \frac{1 -\sqrt{\frac{Tc}{Th}}}{2+\sqrt{\frac{Tc}{Th}}}~~~.
\label{eq:sudefic}
\end{equation}
Equation (\ref{eq:sudefic}) leads to the following hierarchy of the engines maximum work
efficiencies:
\begin{equation}
\eta_s ~\le ~\eta_q ~\leq~ \eta_c~~~.
\label{eq:efic}
\end{equation}
Equation (\ref{eq:efic}) leads to the interpretation that when  the engine is constrained by friction its efficiency
is smaller than the endoreversible efficiency where the engine is constrained by heat transport that is smaller than 
the ideal Carnot efficiency. At the limit of $T_c \rightarrow 0$,
$\eta_s = \frac{1}{2}$ and $\eta_q = \eta_c=1$. 

\begin{figure}
\center{\includegraphics[width=0.65\textwidth]{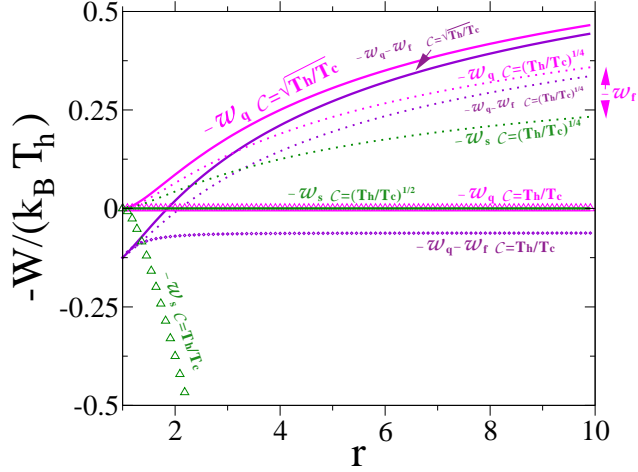}}

\caption{\label{fig:work} Maximum produced work for the quasistatic $-{\cal W}_q$ (magneta), 
quasistatic corrected by friction  $-{\cal W}_q-{\cal W}_f$, when $\frac{\alpha}{\omega_c}=0.5$ (purple) and sudden 
$-{\cal W}_s$ (green) normalized by $k_B T_h$ as a function of the temperature ratio 
$r= \frac{T_h}{T_c}$.
The plot is at the high temperature limit for three compression ratios: ${\cal C}=r$ (triangles) corresponding 
to the Carnot limit, ${\cal C}=r^{1/2}$ (solid) corresponding to the endoreversible optimum
compression ratio and ${\cal C}=r^{1/4}$ (dotted) corresponding to the sudden optimum.
The work against friction in the sudden limit ${\cal W}_f={\cal W}_q-{\cal W}_s$ is indicated.}
\end{figure}

An upper limit to the work invested in friction ${\cal W}_f$ is obtained by subtracting the 
maximum work in the quasistatic limit (\ref{eq:quasistatic work}) from the maximum work in the 
sudden limit (\ref{eq:sudwork1}).
In both these cases infinite heat conductance is assumed leading to $N^B=N_{eq}^h$ and $N^D=N_{eq}^c$. Then 
the upper limit of work invested to counter friction becomes: 
\begin{equation}
{\cal W}_f ~~=~~ \hbar \omega_h \frac{({\cal C}-1)^2 (1+{\cal C}+2{\cal C} N_{eq}^c+2N_{eq}^h)}{4{\cal C}^2}
~~~.
\label{eq:wfric3}
\end{equation}
At high temperature (\ref{eq:wfric3}) simplifies to:
\begin{equation}
{\cal W}_f ~~=~~\frac{1}{2}k_b T_h ({\cal C}-1)^2 ({\cal C}^{-2} + \frac{T_c}{T_h})
~~~. 
\label{eq:wfrichigh}
\end{equation}
Figure \ref{fig:work} compares the maximum produced work at the high temperature limit of
the quasistatic and sudden limits for the compression ratio that optimizes the
quasistatic work ${\cal C}= \sqrt{\frac{T_h}{T_c}}$ and 
the sudden work ${\cal C}= (\frac{T_h}{T_c})^{1/4}$.
The work against friction ${\cal W}_f$ (equation (\ref{eq:wfrichigh})) can be identified the figure as an increasing function of the temperature ratio.
For the compression ratio that optimizes the quasistatic limit the sudden work is zero.
This means that all the useful work is balanced by the work against friction ${\cal W}_f={\cal W}_q$.
Zero produced work is obtained in the quasistatic limit for the compression ratio ${\cal C}=\frac{T_h}{T_c}$ 
and becomes negative for the sudden limit. 

\subsection{Optimal Power Production}

The full range of the performance options of the engine is explored by finding the limit cycle
numerically for a random set of time allocations $\tau_h, \tau_{hc},\tau_c,\tau_{ch}$, 
on the different branches. 

\begin{figure}
\center{\includegraphics[width=0.85\textwidth]{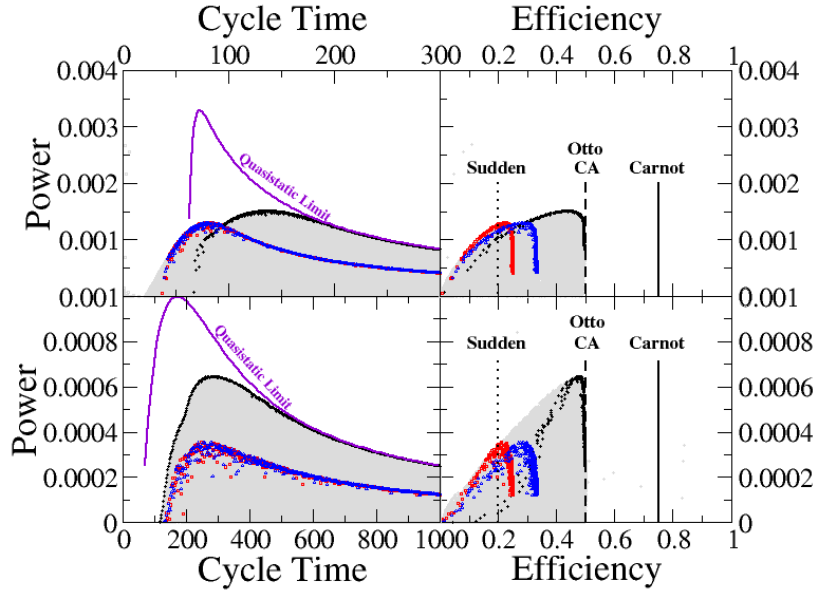}}

\caption{The Power production as a function of the total cycle time (left panels)
and as a function of efficiency (right panels)
for the high temperature limit $\frac{\hbar \omega_h}{k_b T_h}=0.05$
and $\frac{\hbar \omega_c}{k_b T_c}=0.1$ ($\hbar \omega_c=0.025$). The compression ratio is chosen to 
correspond to the optimal value in the quasistatic limit ${\cal C}=\sqrt{{T_h}/{T_c}}$.
Upper panels are for large values of heat transport
($\Gamma=0.6$) and the lower panels are for small value of heat transport ($\Gamma=0.03$).
The graph was produced from a random sample of
time allocations on all branches (gray filled squares). The black circles represent the optimal time 
allocation values which correlate to the quasistatic limit. The blue triangles  
classify the sudden limit on the cold-to-hot {\em adiabat}.
The red squares classify the sudden limit on the hot-to-cold {\em adiabat}.
The Carnot efficiency $\eta_{c}$, endoreversible (CA) efficiency $\eta_{q}$ the Otto efficiency
$\eta_o$ and the sudden efficiency $\eta_{s}$ are indicated in the right panels. The purple line
represents ${\cal P}_q$, the quasistatic power limit  (\ref{eq:quasistaticpower}).}
\label{fig:powerallA}
\end{figure}

Figure \ref{fig:powerallA}
displays the power as a function of total cycle time and efficiency
for the compression ratio ${\cal C}=\sqrt{T_h/T_c}$ that optimizes the 
quasistatic limit at high temperatures.
All plots show a clear global maximum in efficiency at a specific cycle time. In all cases, the deviation
of the maximum power point from the quasistatic estimate is not large. Therefore the 
efficiency at maximum power is quite close to the maximum Otto efficiency $\eta_o$
At large cycle times the optimal power obtained by optimizing the time allocations 
fits very nicely the quasistatic limit.
A closer examination of the cycles with optimal time allocation shows that they form a discontinuous set.
At different total cycle times $\tau$ different combinations of time allocation lead to maximum power.
Underneath the optimal time allocation we can identify cycles which are classified as being sudden
either on the hot-to-cold {\em adiabat} or cold-to-hot {\em adiabat}. 
When the heat conductance $\Gamma$ is high these sudden cycles outperform the quasistatic ones
at short times. Cycles which could be classified as sudden on both {\em adiabats } are absent in 
this plot since they produce almost zero power. This is expected for the compression ratio ${\cal C}= \sqrt{T_h/T_c}$
for which ${\cal W}_s$ is zero (Cf. equation (\ref{eq:wfrichigh}) ).

\begin{figure}
\center{\includegraphics[width=0.85\textwidth]{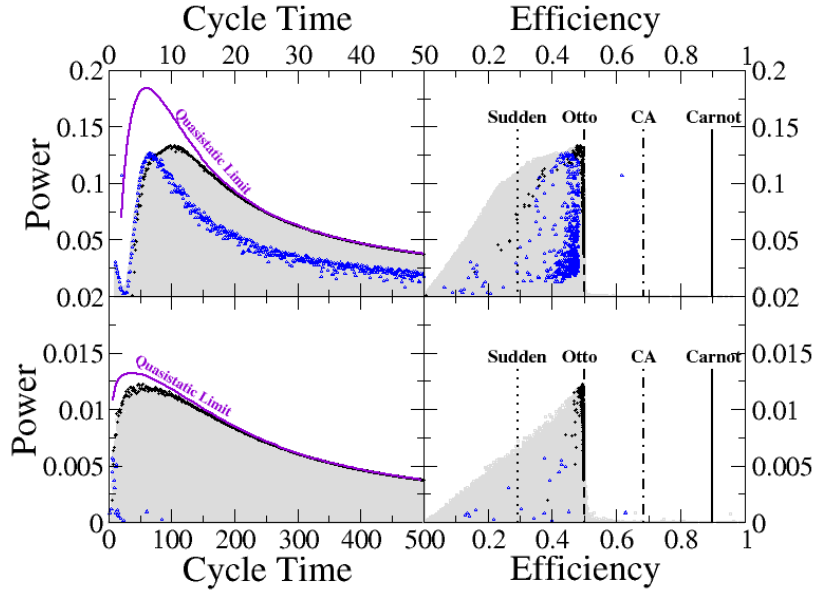}}

\caption{The Power production as a function of the total cycle time (left panels)
and as a function of efficiency (right panels)
for the intermediate temperature limit $\frac{\hbar \omega_h}{k_b T_h}=0.4$
and $\frac{\hbar \omega_c}{k_b T_c}=2.0$ ($\hbar \omega_c=1$). The compression ratio is chosen to 
be  ${\cal C}=(\frac{T_h}{T_c})^{0.3}$.
Upper panels are for large values of heat transport
($\Gamma=0.6$) and the lower panels are for small value of heat transport ($\Gamma=0.03$).
The graph was produced from a random sample of
time allocations on all branches (gray filled squares). The black circles represent the optimal time 
allocation values which correlate to the quasistatic limit. 
The blue triangles  classify time allocations with a shorter time on the {\em isochore} 
relative to the black circle case.
The Carnot efficiency $\eta_{c}$, endoreversible (CA) efficiency $\eta_{q}$ the Otto efficiency
$\eta_o$ and the sudden efficiency $\eta_{s}$ are indicated in the right panels. The purple line
represents the quasistatic power limit (\ref{eq:quasistaticpower}).}
\label{fig:powerallB}
\end{figure}

Figure \ref{fig:powerallB} displays the power as a function of total cycle time and efficiency
for intermediate temperature and a lower compression ratio of ${\cal C}=(\frac{T_h}{T_c})^{0.3}$.
Comparing cycles operating at high temperature (figure \ref{fig:powerallA})
to ones at intermediate temperature (figure \ref{fig:powerallB}) show significant similarities.
The performance of the engine approaches the quasistatic limit when the heat conductance is small.
A careful examination of figure \ref{fig:powerallB} show some cycles with low power output that exceed
the Otto efficiency. This is not a contradiction since the limit restricts only
quasistatic cycles. Thus some sudden like cycles can exceed this limit.
In addition at high heat conductivity a reminiscent of some bang-bang cycles at very
short cycle times can be identified.

\begin{figure}
\center{\includegraphics[width=0.85\textwidth]{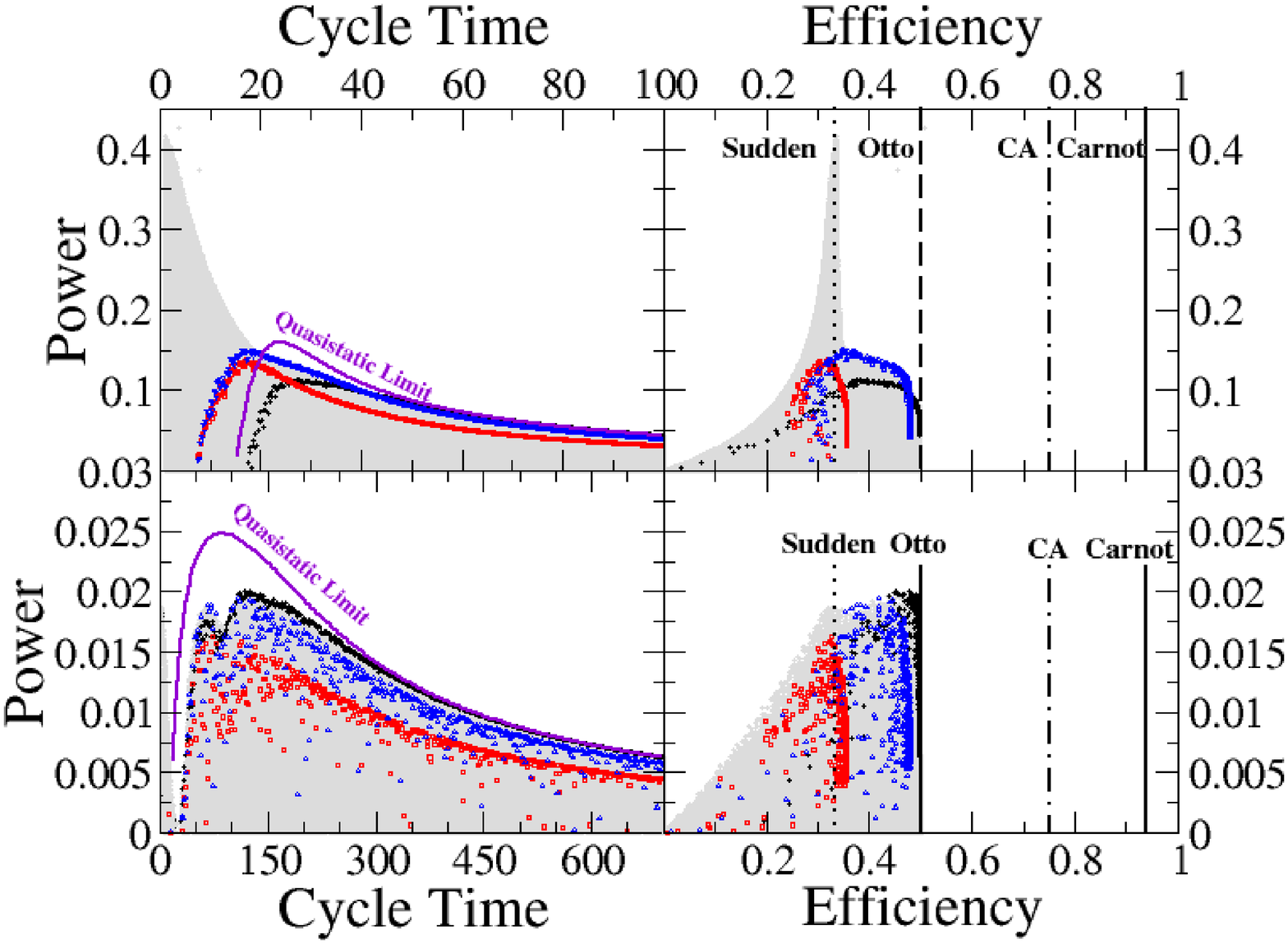}}

\caption{The Power production as a function of the total cycle time (left panels)
and as a function of efficiency (right panels)
for the high temperature limit $\frac{\hbar \omega_h}{k_b T_h}=0.02$
and $\frac{\hbar \omega_c}{k_b T_c}=0.16$ ($\hbar \omega_c=0.1$). The compression ratio is chosen to 
correspond to the optimal value in the sudden limit ${\cal C}=\left(\frac{T_h}{T_c}\right)^{1/4}$.
Upper panels are for large values of heat transport
($\Gamma=0.6$) and the lower panels are for small value of heat transport ($\Gamma=0.03$).
The graph was produced from a random sample of
time allocations on all branches (gray filled squares). The black circles represent the optimal time 
allocation values which correlate to the quasistatic limit. 
The blue triangles  
classify the sudden limit on the cold-to-hot {\em adiabat}.
The red squares classify the sudden limit on the hot-to-cold {\em adiabat}.
The Carnot efficiency $\eta_{c}$, endoreversible (CA) efficiency $\eta_{q}$ the Otto efficiency
$\eta_o$ and the sudden efficiency $\eta_{s}$ are indicated in the right panels. The purple line
represents the quasistatic power limit  (\ref{eq:quasistaticpower}).}
\label{fig:powerallC}
\end{figure}

Operating the engine in the compression ratio corresponding to the  optimum for the 
sudden limit ${\cal C}=(T_h/T_c)^{1/4}$ unravels a different set of optimal time allocations
leading to maximum power. This can be observed in figure \ref{fig:powerallC} which shows
the power as a function of cycle time and its efficiency for this compression ratio.
For conditions of high heat conductance the sudden cycles on both {\em adiabats} out-perform 
all other cycles. These cycles are very close to the bang-bang solutions.
The efficiency at maximum power also corresponds to this limit (Cf. equation (\ref{eq:sudefic})).

\section{Discussion}
\label{sec:discuss}

The performance of the model Otto engine shows close resemblance to macroscopic real engines.
The power has a definite maximum which is the result of a compromise between minimizing
heat transport losses and frictional losses \cite{rebhan02}. How does this phenomena emerge
from the quantum origins of the engine? Put more boldly {\em what is quantum in the engine}?

The quantum laws governing the heat transport from the baths to the working medium (equations (\ref{eq:dissipative}) and
(\ref{eq:hieniso})) have been derived from first principles \cite{breuer,louisell} in the Markovian limit. This limit guarantees the tensor product form of the extended system, allowing for a separation between the local variables of the system and the rest. This separation is a prerequisite for a thermodynamic description \cite{k114}. Nonmarkovian models may also lead to a similar structure \cite {lindblad96}, but are beyond the scope of the present study.
The irreversibility in these equations can be interpreted as emerging from a quantum measurement 
\cite{breuer,knight98,breuer2003}. Specifically for the present system they coincide with the linear Newtonian
heat transport law at high temperature.

The quantum nature of the heat transport on the {\em isochores} becomes
significant at low temperatures when $\frac{\hbar \omega}{k_B T} \ge 1$. 
Planck's constant relates the frequency to a discrete energy unit $ \hbar \omega $. 
The expressions for work (equations (\ref{eq:quasistatic work}) and (\ref{eq:sudwork1})) and 
power production (equation (\ref{eq:quasistaticpower})) are linearly related to  this Quant of energy. 
At higher temperatures the energy unit is replaced by $k_B T$ and $\hbar$ disappears. 
Moreover the thermodynamical expressions  for maximum power and efficiency coincide
with well established relations known from finite time thermodynamics \cite{salamon01,broeck05}. 
The finding is in line with the study of Geva et. al. \cite{k87} which has shown that although the
quantum heat transport laws are not linear in general (equation (\ref{eq:hettrans})), and the system 
can be far from equilibrium, nevertheless they lead at high temperatures to the 
Curzon Ahlborn efficiency \cite{curzon75} at maximum power.

The rapid dynamics on the {\em adiabatic} branches is responsible for the frictional behaviour. 
The quantum origin can be traced to the noncommutability of the Hamiltonian at different times.
As a result the system is not able to  adiabatically follow the instantaneous energy 
representation. 
For the harmonic model the origin of the noncommutability comes about from
the kinetic energy not commuting with the potential energy, the origin of which
is the fundamental relation $[\Op Q, \Op P] = i h$. Plank's constant appearing
in the commutation relation of the kinetic and potential energy 
$[ \frac{\Op P^2}{2m}, \frac{k \Op Q^2}{2}]=i\hbar \frac{\omega^2}{2} \Op D$ (Cf. equation (\ref{eq:hamils})) is cancelled by the inverse of Planck's constant appearing in the equation of motion
(\ref{eq:lvn}) or (\ref{eq:hieniso}). 

The deviation from adiabaticity is independent of the mass of the oscillator.
This inertial effect appears only in the combination $\alpha = \dot \omega / \omega $ 
the nonadiabatic parameter. This means that friction like behaviour is expected whenever
the internal timescale of the working medium is comparable to the engines cycle time. Without a dissipative branch, however, this ``friction'' cannot be interpreted as such - it is even reversible.

The most important noncommutability of the engine is between the branch propagators
$[{\cal U}_{ch},{\cal U}_{c}] \neq 0$ or $[{\cal U}_{hc},{\cal U}_{h}] \neq 0$, 
composing the cycle propagator ${\cal U}_{cyc}$ in equation (\ref{eq:cycprop}). 
The sequence of alternating propagations of {\em adiabats} and {\em isochores} 
is what breaks the time reversal symmetry
and allows the engine to produce power irreversibly.

\section{Conclusion}
\label{sec:conclusion}

The quantum Otto cycle model engine served its purpose of unravelling the 
emergence of thermodynamics out of quantum mechanics. Due to the simplicity of
a working medium composed of harmonic oscillators the following insights have been illustrated:
\begin{itemize}
\item{Thermodynamical variables are related to quantum observables which are
sufficient to reconstruct the state of the working medium. The energy variable is supplemented by
additional variables required to describe the deviation from adiabatic behaviour in the limit cycle.
The knowledge of these observables enables the calculation of thermodynamical variables
such as entropy and temperature. The first law emerges as a statement accounting for the energy change 
in an open quantum system. The second law is related to the fact that the energy entropy
is always larger than the von Neumann entropy which is invariant to free evolution.
The information required to restore the von Neumann entropy from the energy entropy is partially erased
on the {\em isochores}. The open dynamics on the {\em isochores} can be considered as a partial measurement
of energy. As a result information on the operators $\Op L$ and $\Op D$ which do not commute with energy is lost.}

\item{Quantum origins of irreversible losses are due to heat transport and with it the unavoidable
dephasing. Quasistatic cycles which almost follow adiabatically the energy frame
show irreversible losses dominated by heat transport. The optimal performance of
these cycles closely resembles the endoreversible limit.} 

\item{Deviation from quantum adiabatic behaviour is reflected by
frictional losses which emerge from the generation of inertial components on
the {\em adiabats} and their dephasing on the {\em isochores}. The frictional loss terms
are independent of the cycle direction of motion.}

\item{At high temperatures, explicit quantum variables and constants disappear. The performance of the engine shows
universal behaviour which is a function of temperature $T$ compression ratio ${\cal C}$
and the ratio between the cycle frequency proportional to $\alpha$ and the internal frequency 
of the working medium $\omega$.}

\end{itemize}

\ack
We want to thank Tova Feldmann and Maurice Cohen for their  help and Jeffrey Gordon and David Tannor
for stimulating discussions. This work was supported
by the Israel Science Foundation, The Fritz Haber centre is supported by the Minerva
Gesellschaft f\"ur die Forschung, GmbH M\"unchen, Germany.

\appendix
In addition to the derivations outlined below, a full derivation of the results presented in this paper can be found in the Mathematica notebook file, ``HOE.nb'' which can be downloaded from www.fh.huji.ac.il/\char126 yairr.
\section{Canonical invariance for the product form}
\label{ap:canonical}
The canonical invariance of a density operator represented as a product of exponents is considered:
\begin{equation}
\Op \rho ~~=~~  e^{ \mu_1 \Op X_1}e^{ \mu_2 \Op X_2} ...e^{ \mu_k \Op X_k}...e^{ \mu_N \Op X_N}
\label{eq:prd1}
\end{equation}
where the set $\{ \Op X \}$ form a closed Lie algebra and $\mu=\mu(t)$ 
are time dependent coefficients. 
Canonical invariance means that the form of (\ref{eq:prd1}) is preserved during the evolution.
To test this hypothesis it can be noted that $\dot {\Op \rho \Op {\rho^{-1}}}$ 
forms a linear combination of the set of operators $\{ \Op X \}$:
\begin{equation}
\frac{d \Op \rho}{dt}\Op \rho^{-1}~~=~~ \dot \mu_1 \Op X_1 ~~+~~ 
\dot \mu_2  e^{ \mu_1 \Op X_1} {\Op X_2}e^{- \mu_1 \Op X_1}~~+~~ 
\dot \mu_3 e^{ \mu_1 \Op X_1}  e^{ \mu_1 \Op X_2}{\Op X_3}
e^{- \mu_1 \Op X_2}e^{- \mu_1 \Op X_1} ~+~ ...
\label{eq:prod2}
\end{equation}
The Baker- Hausdorff relation $e^{\Op A} {\Op B}e^{-\Op A}~~=~~ {\Op B}~+~
[{\Op A},{\Op B}]~+~\frac{1}{2\!}[{\Op A},[{\Op A},{\Op B}]] ...$ and the closure
of the Lie algebra to commutation means that the RHS of equation (\ref{eq:prod2}) is 
a linear combination of the operators $\{ \Op X \}$ in the algebra.

For canonical invariance to prevail during the evolution 
${\cal L}({\Op \rho}){\Op \rho}^{-1}$ also should become a linear combination of the set
of operators $\{ \Op X \}$. This will lead to $N$ coupled differential
equations for the coefficients $\mu(t)$.
 
The simpler case is when the Hamiltonian is also composed of a linear combination
of the same algebra then 
${\cal L}({\Op \rho}){\Op \rho}^{-1}= -i \hbar [{\Op H},{\Op \rho}] {\Op \rho}^{-1}
= -i \hbar{\Op H}-i \hbar {\Op \rho}{\Op H} {\Op \rho}^{-1} $
forming a linear sum of the same set  $\{ \Op X \}$ and canonical
invariance prevails. 

For an open quantum system the equivalent statement would be that in addition
${\cal L}_D({\Op \rho}) {\Op \rho}^{-1}$ also forms a linear
combination of the set $\{ \Op X \}$.  This condition becomes:
\begin{equation}
{\cal L}_D ({\Op \rho}){\Op \rho}^{-1}~~=~~ 
\sum_j ~\left( {\Op F_j }{\Op \rho} \Op  F_j^{\dagger} 
{\Op \rho}^{-1}
-\frac{1}{2} (\Op F_j^{\dagger}  \Op F_j+ {\Op \rho} {\Op F_j} {\Op F_j}^{\dagger} {\Op \rho}^{-1}
) \right)
\label{eq:canonical3}
\end{equation}
where $\Op F$ are Lindblad operators \cite{lindblad76}. Equation (\ref{eq:canonical3}) will form 
a linear combination of the set of operators $\{ \Op X \}$ if 
the operators $\Op F_j  \Op F_j^{\dagger}$ and 
${\Op F_j }{\Op \rho} \Op  F_j^{\dagger} {\Op \rho}^{-1} $
are part of the set $\{ \Op X \}$ . 

Specifically for the harmonic oscillator set 
$\Op P^2$ , $\Op Q^2$ and $\Op D =  (\Op Q \Op P + \Op P \Op Q)$ and the Lindblad operator
$\Op F =\Op a$ this condition is met. On the contrary for $\Op F = \Op H$ , $\Op H^2$ is not
part of the Lie algebra therefore, canonical invariance is not met in the case of pure dephasing.

\section{The relations between the parameters in the product form and the exponential form}
\label{sec:relation}

The state of the system can be expressed in an exponent product form \eref{eq:product}
or as an exponent of an operator sum \eref{eq:stateexp}. 
To determine this equivalence, the expression  $\Op \rho \Op a \Op \rho^{-1}$ and
$\Op \rho \Op a^{\dagger}\Op \rho^{-1}$ are evaluated. 
The normalization $Z$ is identical for both forms and therefore is not considered. In the exponent-product
form \eref{eq:product} using the  Baiker-Housdorff formula one obtains:
\begin{equation}
\Op {\rho}\Op {a}\Op {\rho}^{-1}=(e^{\beta\hbar\omega}-4e^{-\beta\hbar\omega}\gamma\gamma^{*})
\Op {a}-2e^{-\beta\hbar\omega}\gamma^{*}\Op {a^{\dagger}}
\label{eq:appb1}
\end{equation}

\begin{equation}
\Op {\rho}\Op {a}^{\dagger}\Op{\rho}^{-1}=2e^{-\beta\hbar\omega}\gamma\Op{a}+e^{-\beta\hbar\omega}\Op{a^{\dagger}}
\end{equation}
Using the operator sum form \eref{eq:stateexp} to evaluate the same two expressions,
leads to:
\begin{eqnarray}
\nonumber
\Op {\rho}\Op {a}\Op{\rho}^{-1} & = & \left(\cosh(\sqrt{\left(\hbar\omega\chi_{2}\right)^{2}-4\chi_{1}
\chi_{1}^{*}})-\chi_{2}\frac{\hbar\omega}{\sqrt{\left(\hbar\omega\chi_{2}\right)^{2}-4\chi_{1}\chi_{1}^{*}}}
\sinh(\sqrt{\left(\hbar\omega\chi_{2}\right)^{2}-4\chi_{1}\chi_{1}^{*}})\right)\Op {a}\\
 &  & -\left( \frac{2\chi_{1}^{*}}{\sqrt{\left(\hbar\omega\chi_{2}\right)^{2}-4\chi_{1}
\chi_{1}^{*}}}\sinh(\sqrt{\left(\hbar\omega\chi_{2}\right)^{2}-4\chi_{1}\chi_{1}^{*}})\right)\Op {a^{\dagger}}
\end{eqnarray}

and
\begin{eqnarray}
\nonumber
\Op {\rho}\Op {a^{\dagger}}\Op {\rho}^{-1} & = & \left(\cosh(\sqrt{\left(\hbar\omega\chi_{2}\right)^{2}-4\chi_{1}
\chi_{1}^{*}})+\chi_{2}\frac{\hbar\omega}{\sqrt{\left(\hbar\omega\chi_{2}\right)^{2}-4\chi_{1}
\chi_{1}^{*}}}\sinh(\sqrt{\left(\hbar\omega\chi_{2}\right)^{2}-4\chi_{1}\chi_{1}^{*}})\right)\Op {a^{\dagger}}\\
 &  & +\left( \frac{2\chi_{1}}{\sqrt{\left(\hbar\omega\chi_{2}\right)^{2}-4\chi_{1}
\chi_{1}^{*}}}\sinh(\sqrt{\left(\hbar\omega\chi_{2}\right)^{2}-4\chi_{1}\chi_{1}^{*}})\right) \Op {a}
\label{eq:appb2}
\end{eqnarray}
Equating both formula pairs results in the identification:
\begin{eqnarray}
\label{ea:relation}
\chi_{1}=2\frac{\mathrm{arcsinh}\left(\frac{1}{2}\gamma 
e^{-\beta\hbar\omega}\sqrt{\frac{-16\gamma\gamma^{*}
+\left(e^{2\beta\hbar\omega}-1-4\gamma\gamma^{*}\right)^{2}}{\gamma^{2}}}\right)}
{\sqrt{\frac{-16\gamma\gamma^{*}+\left(e^{2\beta\hbar\omega}-1-4\gamma\gamma^{*}\right)^{2}}
{\gamma^{2}}}}
\\
\nonumber
\chi_{1}^{*}=\chi_{1}\frac{\gamma^{*}}{\gamma}
~~~~~~,~~~~~~
\chi_{2}=\chi_{1}\frac{1-e^{2\beta\hbar\omega}+4\gamma\gamma^{*}}{2 \gamma \hbar \omega}
\end{eqnarray}
which establishes the relation between  $\gamma$ and $\beta$ in the product form \eref{eq:stateexp}
and $\chi$  in the the operator sum form \eref{eq:product}.
\pagebreak

\section*{References}


\end{document}